\documentclass[10pt, conference, compsocconf]{IEEEtran}
\usepackage{graphicx}
\usepackage[cmex10]{amsmath}
\usepackage{amsthm,amssymb}
\usepackage{lineno}

\theoremstyle{definition}
\newtheorem{thm}{Theorem}[section]

\usepackage{url}
\usepackage{subfig}

\usepackage{color}
\usepackage[colorlinks=true,citecolor=blue,urlcolor=blue]{hyperref}

\newcommand{\dt}{\Delta t}
\newcommand{\dx}{\Delta x}

\newcommand*\patchAmsMathEnvironmentForLineno[1]{%
  \expandafter\let\csname old#1\expandafter\endcsname\csname #1\endcsname
  \expandafter\let\csname oldend#1\expandafter\endcsname\csname end#1\endcsname
  \renewenvironment{#1}%
     {\linenomath\csname old#1\endcsname}%
     {\csname oldend#1\endcsname\endlinenomath}}%
\newcommand*\patchBothAmsMathEnvironmentsForLineno[1]{%
  \patchAmsMathEnvironmentForLineno{#1}%
  \patchAmsMathEnvironmentForLineno{#1*}}%
\AtBeginDocument{%
\patchBothAmsMathEnvironmentsForLineno{equation}%
\patchBothAmsMathEnvironmentsForLineno{align}%
\patchBothAmsMathEnvironmentsForLineno{flalign}%
\patchBothAmsMathEnvironmentsForLineno{alignat}%
\patchBothAmsMathEnvironmentsForLineno{gather}%
\patchBothAmsMathEnvironmentsForLineno{multline}%
}

\begin{document}

\title{Parallel Semi-Implicit Time Integrators}

\author{
  \IEEEauthorblockN{Benjamin Ong}
  \IEEEauthorblockA{Michigan State University\\
    Institute for Cyber Enabled Research\\
    East Lansing, MI USA \\
    Email: ongbw@msu.edu}
  \and
  \IEEEauthorblockN{Andrew Melfi}
  \IEEEauthorblockA{Michigan State University\\
    East Lansing, MI USA \\
    Email: melfiand@msu.edu}
  \and
  \IEEEauthorblockN{Andrew Christlieb}
  \IEEEauthorblockA{Michigan State University\\
    Department of Mathematics\\
    East Lansing, MI USA \\
    Email: andrewch@math.msu.edu}
}

\maketitle

  \begin{abstract}
    In this paper, we further develop a family of parallel time
    integrators known as Revisionist Integral Deferred Correction
    methods (RIDC) to allow for the semi-implicit solution of time
    dependent PDEs.  Additionally, we show that our semi-implicit RIDC
    algorithm can harness the computational potential of {\em
      multiple} general purpose graphical processing units (GPUs) in a
    single node by utilizing existing CUBLAS libraries for matrix
    linear algebra routines in our implementation.  In the numerical
    experiments, we show that our implementation computes a fourth
    order solution using four GPUs and four CPUs in approximately the
    same wall clock time as a first order solution computed using a
    single GPU and a single CPU.

  \end{abstract}


\section{Introduction}

RIDC methods are parallel--in--step time integrators
\cite{ridc,implicit-ridc}.  The ``revisionist'' terminology was first
adopted in \cite{ridc} to highlight that (i) this is a \emph{revision}
of the standard integral (or spectral) defect correction (IDC or SDC)
methods \cite{IDC1,IDC2,idc-ark, DGR, layton2007icp,minion2003}, and
(ii) successive corrections, running in parallel but lagging in time,
\emph{revise} and improve the approximation to the solution.  This
notion of time parallelization is particularly exciting because it can
be potentially layered upon existing spatial parallelization
techniques \cite{ridc-dd,TW05}, including algorithms that utilize GPU
cards to solve time dependent PDEs \cite{jk11,pde-gpu,ALT08}, to add
further parallel scalability.


The main idea behind RIDC methods is to re-write the defect correction
framework \cite{defect1,defect2} so that, after initial start-up
costs, each correction loop can be lagged behind the previous
correction loop in a manner that facilitates running the predictor and
correctors in parallel.  This idea for parallel time integrators was
previously published by the present authors in
\cite{ridc,implicit-ridc}.  As before, this is still small scale
parallelism in the sense that the time parallelization is limited by
the order one wants to achieve.

To harness the computational potential of {\em multiple} graphical
processing units (GPUs) on a single node, the CUBLAS library
\cite{cuda} (which is a collection of linear algebra subroutines coded
in CUDA) are utilized to demonstrate that by threading the RIDC loops,
multiple GPUs can be utilized for our semi-implicit RIDC algorithm.
We present numerical experiments in Section~\ref{sec:benchmarks} to
show that our algorithm and implementation computes a fourth order
semi-implicit solution using four GPUs and four CPUs in approximately
the same wall clock time as a first order forward-backward Euler
solution computed using a single GPU and a single CPU.  We stress that
our parallel speedup comes from a unique way to utilize existing
parallel libraries, in this case the CUBLAS libraries provided by
NVIDIA.  Unless data decomposition is used (whether in the host code
or within a new parallel library), one could not use multiple GPU
cards in as simple a fashion using a sequential ARK integrator.  We
believe that this work will provide the scientific community with a
straightforward way to add further parallelism to existing software
that generate low order (in time) solutions to time dependent PDEs
using only a single GPU card.

Readers might be familiar with parareal integrators
\cite{ganderhairer08, gandervandewalle07, lionsmadayturinici01,
  madayturinici02, gander07}, another family of parallel time
integrators.  In such methods, the time domain is split into
sub-problems that can be computed in parallel, and an iterative
procedure for coupling the sub-problems is applied, so that the
overall method converges to the solution of the full problem.
Parareal integrators are philosophically different from RIDC methods.
While parareal methods allow for large scale parallelization, there
are non trivial choices of the fine and coarse predictor that affect
convergence to the desired solution.  RIDC methods on the other hand,
guarantees convergence, and high order solutions.  A class of parareal
methods by Mike Minion and Matthew Emmett (CAMCOS, 2012) are
potentially a generalization of the RIDC methods, but further analysis
would be needed to validate that statement.


This paper is organized as follows: In Section~\ref{sec:ark}, we
review Implicit--Explicit (IMEX) methods, which are a family of high
order semi-implicit integrators \cite{imex,kencar}.  In Section~\ref{sec:RIDC},
semi-implicit RIDC methods and their properties are presented.  Then,
numerical benchmarks comparing RIDC and additive Runge--Kutta methods
are given in Section~\ref{sec:benchmarks}, followed by concluding
remarks in Section~\ref{sec:conclusion}.

\section{IMEX Methods}
\label{sec:ark}
We are interested in solutions to initial value problems of the form,
\begin{align}
  \left\{\begin{array} {l}
      \displaystyle
      y'(t)  = f^S(t,y) + f^N(t,y), \quad t \in[a, b], \\
      \displaystyle
      y(a) = \alpha.
    \end{array}
  \right.
  \label{eqn:ode}
\end{align}
where $y,\alpha\in\mathbb{R}^n$,
$f^N:\mathbb{R}\times\mathbb{R}^n\to\mathbb{R}^n$ and
$f^S:\mathbb{R}\times\mathbb{R}^n\to\mathbb{R}^n$.  The function
$f^S(t,y)$ contains stiff terms that need to be handled implicitly,
and $f^N(t,y)$ consists of non-stiff terms that can be handled
explicitly.  A first order implicit-explicit (IMEX) discretization of
the IVP~\eqref{eqn:ode} can be written as
\begin{align*}
  \frac{y_{n+1}-y_{n}}{\Delta t}  = f^S(t_{n+1},y_{n+1}) 
  + f^N(t_n,y_n),  \text{ with }
  y_0 = \alpha.
\end{align*}
The above IMEX discretization is particularly useful if $f^S(t,y)$ is
linear in $y$, i.e. $f^S(t,y) = Dy$, which is often the case in a
method of lines discretization of PDEs containing relaxation terms.
In such cases, the IMEX discretization reduces to a linear system
solve the solution at each time level,
\begin{align}
  (I - D\Delta t)y_{n+1} = y_{n}
  + \Delta t f^N(t_n,y_n),  \quad \text{with }
  y_0 = \alpha.
  \label{eqn:imex_diffusion}
\end{align}
An $s$-stage diagonally implicit RK (DIRK) and explicit $s$-stage
explicit RK method are coupled
\begin{align*}
  \renewcommand{\arraystretch}{1.5}
  \begin{array}{c|cccc|ccccc}
    0 & 0 &  &  & & 0 & & & &\\
    c_2 & a_{21}^{S} & a_{22}^{S} & 0 & \ldots & a_{21}^{N} & 0 & & &\\ 
    \vdots & \vdots & \vdots &\ddots & & \vdots & & & \ddots &\\
    c_s & a_{s1}^{S} & a_{s2}^{S}  & \cdots  & a_{ss}^{S}  & 
    a_{s1}^{N}  & a_{s2}^{N} & \cdots   & a_{s,s-1}^{N}  &  0  \\  
    \hline
    & b_{1}^{S} & b_{2}^{S} & \cdots & b_{s}^{S} & 
    b_{1}^{N} & b_{2}^{N} & \cdots & b_{s-1}^N & b_{s}^{N} 
  \end{array}
\end{align*}
to generate a high order semi-implicit integrator.  The discretization
of IVP~\eqref{eqn:ode} using an IMEX method can be written as
\begin{align*}
  y_{n+1}  = y_n + \Delta t \sum_{i=1}^s \left(b_i^SK_{ni}^S
  + b_i^NK_{ni}^N\right),  \quad \text{with }
  y_0 = \alpha,
\end{align*}
where the stages satisfy
\begin{align*}
  K_{ni}^S = f^S(t_{ni},\, y_n + 
  \Delta t \sum_{j=1}^{i}a_{ij}^SK_{nj}^S
  + \Delta t \sum_{j=1}^{i-1} a_{ij}^N K_{nj}^N)\\
  K_{ni}^N = f^N(t_{ni},\, y_n + 
  \Delta t \sum_{j=1}^{i}a_{ij}^SK_{nj}^S
  + \Delta t \sum_{j=1}^{i-1} a_{ij}^N K_{nj}^N).
\end{align*}
with $t_{ni}=t + c_i\Delta t$.  The third and fourth order IMEX
methods from \cite{liuzou} are used to benchmark against our fourth
order RIDC-FBE (RIDC constructed using forward and backward Euler
integrators).  The third order IMEX method is constructed from the
following Butcher tableaux:
\begin{align*}
 \renewcommand{\arraystretch}{1.3}
  \begin{array}{c|rrrrr|rrrrr}
    0 & 0 & & & & & 0 &&& & \\
    \frac12 & 0 & \frac12 & && &
    \frac12 &&&&\\
    \frac12 & \frac14 & -\frac5{12} & \frac23&& &
    \frac14 & \frac14 &&&\\
    1 & 2 & -\frac72  & \frac12 & 2  & &
    0 & 1 & 0 &&\\
    1 & \frac16 & 0 & \frac23 & -\frac56 & 1  &
    \frac16 & 0 & \frac23 & \frac16 &\\
    \hline
     & \frac16 & 0 & \frac23 & -\frac56 & 1  &
    \frac16 & 0 & \frac23 & \frac16 &
  \end{array}
\end{align*}
 The fourth order IMEX method is constructed from the following
DIRK method
\begin{align*}
 \renewcommand{\arraystretch}{1.3}
  \begin{array}{c|rrrrrrr}
    0 & 0 & & & & & & \\
    \frac13 & -\frac16 & \frac12&&&&&  \\
    \frac13 & \frac16 & -\frac13 & \frac12&&&& \\
    \frac12 & \frac38 & -\frac38 & 0 & \frac12 &&& \\
    \frac12 & \frac18 & 0 & \frac38 & -\frac12 & \frac12 && \\
    1 & -\frac12 & 0 & 3 & -3 & 1 & \frac12 & \\
    1 & \frac16 & 0 & 0 & 0 & \frac23 & -\frac12 & \frac23 \\
    \hline
     & \frac16 & 0 & 0 & 0 & \frac23 & -\frac12 & \frac23 
  \end{array}
\end{align*}
and explicit RK method
\begin{align*}
 \renewcommand{\arraystretch}{1.3}
  \begin{array}{c|rrrrrrr}
    0  
    & 0 &&&&&&\\
    \frac13 
    & \frac13 &&&&&&\\
    \frac13 
    & \frac16 & \frac16 &&&&&\\
    \frac12 
    & \frac18 & 0 & \frac38 &&&&\\
    \frac12 
    & \frac18 & 0 & \frac38 & 0 &&&\\
    1 
    & \frac12 & 0 & -\frac32 & 0 & 2 &&\\
    1 
    & \frac16 & 0 & 0 & 0 & \frac23 & \frac16 &\\
    \hline
    & \frac16 & 0 & 0 & 0 & \frac23 & \frac16 & 0
  \end{array}
\end{align*}
(Note, a second order IMEX scheme was also tested, but not presented
because of its poor stability constraints).  Similar to before, if
$f^S(t,y)$ is linear in $y$, then each stage computation reduces to a
linear solve since a DIRK method was paired with an explicit
integrator in the discussed IMEX methods.

\section{RIDC Methods}
\label{sec:RIDC}
\begin{figure}[htbp]
  \centering
  \includegraphics[width=0.45\textwidth]{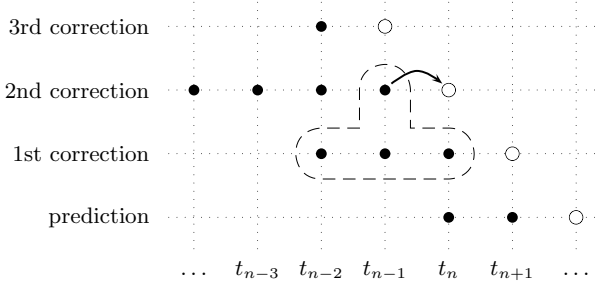}
  \caption{(RIDC4-FBE) This plot shows the staggering required for a
    fourth order RIDC scheme, constructed using a first order
    implicit--explicit predictors and correctors.  The time axis runs
    horizontally, and the correction levels run vertically. The white
    circles denote solution values that are simultaneously computed,
    e.g., core 0 is computing the prediction solution at time
    $t_{n+2}$ while core 1 is computing the 1st corrected solution at
    time $t_{n+1}$, etc.  }
  \label{fig_stencil_ridc4be}
\end{figure}

RIDC methods are a class of time integrators based on integral
deferred correction \cite{DGR00}.  RIDC methods first compute a
prediction to the solution (``level 0'') using low order schemes
(e.g. a first order implicit-explicit method) followed by one or more
corrections to compute subsequent solution levels.  Each correction
\emph{revises} the solution and increases the formal order of accuracy
by $1$, if a first order implicit-explicit integrator is used to solve
the error equation.  Each correction level is delayed from the
previous level as illustrated in Figure~\ref{fig_stencil_ridc4be} --
the open circles denote solution values that are simultaneously
computed. This staggering in time means that the predictor and each
corrector can all be executed simultaneously, in parallel, while each
processes a different time-step.

In Section~\ref{sec:error_eqn}, we first derive the error equation.
Then, Section~\ref{sec:predictor} and Section~\ref{sec:corrector} give
numerical schemes for solving the IVP and the error equation.  In
Section~\ref{sec:order}, we review theorems related to the formal
order of accuracy that follow trivially from \cite{idc-ark}, and in
Section~\ref{sec:startstop}, we summarize starting and stopping
details for the RIDC algorithm as well as the notion of restarts.

\subsection{Error Equation}
\label{sec:error_eqn}
Suppose an approximate solution $\eta(t)$ to IVP~\eqref{eqn:ode} is
computed.  Denote the exact solution as $y(t)$.  Then, the error of
the approximate solution is
\begin{align}
  \label{eqn:error}
  e(t) = y(t)-\eta(t).
\end{align}
If we define the residual as $ \epsilon(t) = \eta'(t) - f^S(t,\eta(t))
- f^N(t,\eta(t))$, then the derivative of the error \eqref{eqn:error}
satisfies
\begin{align*}
  e'(t) &= y'(t) - \eta'(t) \\
  &= f^S(t,y(t))  + f^N(t,y(t)) -
  f^S(t,\eta(t)) \\
  &\phantom{=} - f^N(t,\eta(t)) - \epsilon(t).
\end{align*}
The integral form of the error equation, 
\begin{align}
  \nonumber
  \left[e(t)+\int_a^t\epsilon(\tau)\,d\tau\right]' 
  = f^S\left(t,\eta(t)+e(t)\right)    - f^S\left(t,\eta(t)\right) \\
  \label{eqn:error_eqn}
  + f^N\left(t,\eta(t)+e(t)\right) 
 - f^N\left(t,\eta(t)\right),
\end{align}
can then be solved using the initial condition $e(a) = 0$.

\subsection{The predictor}
\label{sec:predictor}
To generate a provisional solution that can be corrected, a low order
integrator is applied to solve IVP~\eqref{eqn:ode}; this process is
typically known as the prediction loop.  The first-order IMEX scheme
reviewed in Section~\ref{sec:ark} will be used to generate our
RIDC-FBE (RIDC forward and backward Euler method) though in theory,
any IMEX methods reviewed in Section~\ref{sec:ark} can be used.  We
adopt the following notation:
\begin{align}
  \eta^{[0]}_{n+1} = \eta^{[0]}_{n} + \dt_n
  f^S(t_{n+1},\eta^{[0]}_{n+1}) + \dt_n f^N(t_{n},\eta^{[0]}_{n}),
  \label{eqn:pred}
\end{align}
where the superscript $^{[0]}$ indicates this is the solution at level
$0$, the prediction level.  This non-linear equation can be solved
using Newton's method.  

\subsection{The corrector}
\label{sec:corrector}
The correctors are also low order integrators, but are used to solve
the error equation~\eqref{eqn:error_eqn} for the error $e(t)$ to an
approximate solution $\eta(t)$.  Since the error equation is solved
iteratively to improve a solution from the previous level, each
correction level computes an error $e^{[j-1]}(t)$ to the solution at
the previous level $\eta^{[j-1]}(t)$ to obtain a revised solution
$\eta^{[j]}(t) = \eta^{[j-1]}(t) + e^{[j-1]}(t)$.

A first order IMEX discretization of the error
equation~\eqref{eqn:error_eqn} (after some algebra) gives
\begin{align}  
  \label{eqn:corr}
  \eta^{[j]}_{n+1} = \eta^{[j]}_{n} + \Delta t \left[f^S(t_{n+1},
    \eta^{[j]}_{n+1}) + f^N(t_{n}, \eta^{[j]}_{n})\right]  \\
  \nonumber
  -\left[\Delta t f^S(t_{n+1}, \eta^{[j-1]}_{n+1}) + f^N(t_{n},
    \eta^{[j-1]}_{n})\right]  \\
  \nonumber
  + \int_{t_n}^{t_{n+1}}
  f(\tau,\eta^{[j-1]}(\tau))\,d\tau.
\end{align}
The integral $\int_{t_n}^{t_{n+1}}f(\tau,\eta^{[j-1]}(\tau))\,d\tau$
is approximated using quadrature.  For the $j$\textsuperscript{th}
correction loop, $(j+1)$ nodes are needed in the stencil to accurately
approximate the integral.  There are various choices for the stencil,
but in practice, the stencil should include the nodes $t_n$ and
$t_{n+1}$.  We make the following choice for selecting our quadrature nodes:
\begin{align}  
  &\int_{t_n}^{t_{n+1}}f(\tau,\eta^{[j-1]}(\tau))\,d\tau \approx\\
  \nonumber
  &\begin{cases}
    \sum_{k=0}^j \alpha_{nk} \left(f^N(t_{n+1-k},\eta^{[j-1]}_{n+1-k}) 
    + f^S(t_{n+1-k},\eta^{[j-1]}_{n+1-k})\right), \\
    \hspace*{2.5in}\mbox{if } (n \ge j-1)  \\   
    \sum_{k=0}^j \alpha_{nk} \left(f^N(t_{k},\eta^{[j-1]}_{k}) 
    + f^S(t_{k},\eta^{[j-1]}_{k})\right), \\
    \hspace*{2.5in}\mbox{if }  (n<j-1)
  \end{cases}
\end{align}
where the quadrature weights are given by
\begin{align*}
  \alpha_{nk} = \int_{t_n}^{t_{n+1}} \prod_{i=0, i\neq k}^{j}
  \frac{(t-t_{n+1-i})}{(t_{n+1-k}-t_{n+1-i})}\,dt,
\end{align*}
for $n \ge j-1$, $k = 0,1,\ldots,j-1$, and 
\begin{align*}
  \alpha_{nk} = \int_{t_n}^{t_{n+1}} \prod_{i=0, i\neq k}^{j}
  \frac{(t-t_{i})}{(t_{k}-t_{i})}\,dt,
  \quad k = 0,1,\ldots,j-1
\end{align*}
for $ n<j-1$.  Since uniform time steps are used in the computation, then
only one set of quadrature weights needs to be computed, stored, then
used as necessary.

\subsection{Formal order of accuracy}
\label{sec:order}
The analysis in \cite{idc-ark}, proving convergence under mild
conditions for IDC-IMEX methods, extends simply to these RIDC-IMEX
methods.

\begin{thm} 
  Let $f(t,y)$ and $y(t)$ in IVP~\eqref{eqn:ode} be sufficiently
  smooth.  Then, the local truncation error for a RIDC method
  constructed using a first order IMEX integrators
  for the predictor and $(p-1)$  correction loops is $\mathcal{O}
  (\Delta t^{p+1})$.
\end{thm}
\begin{thm} 
  Let $f(t,y)$ and $y(t)$ in IVP~\eqref{eqn:ode} be sufficiently
  smooth.  Then, the local truncation error for an RIDC method
  constructed using uniform time steps, a
  $p_0$\textsuperscript{th}-order ARK method in the
  prediction loop, and
  $(p_1,p_2,\cdots,p_{j})$\textsuperscript{th}-order ARK methods in the
  correction loops, is $\mathcal{O} (\Delta t^{p+1})$, where $p =
  \sum_{i=0}^{j} p_i$.
\end{thm}

\subsection{Further Comments}
\label{sec:startstop}
\begin{figure}[htbp]
  \centering
  \includegraphics[width=0.45\textwidth]{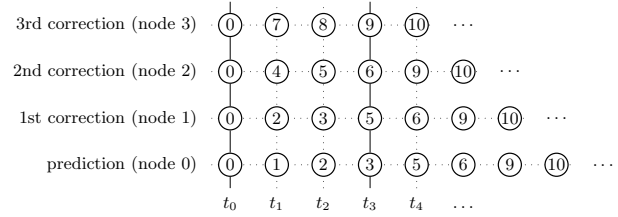}
  \caption{This figure is a graphical representation of how the
    RIDC4-BE algorithm is started.  The time axis runs horizontally,
    the correction levels run vertically.  All nodes are initially
    populated with the initial data at $t_0$.  This is represented by
    computing step 0 (enclosed in a circle).  At computing step 1,
    node 0 computes the predicted solution at $t_1$.  The remaining
    nodes remain idle.  At computing step 2, node 0 computes the
    predicted solution at time $t_2$, node 1 computes the 1st
    corrected solution at time $t_1$, the remaining two nodes remain
    idle.  Note that in this starting algorithm, special care is taken
    to ensure that minimum memory is used by not letting the computing
    cores run ahead until they can be marched in a pipeline; in this
    example, when node 3 starts computing $t_3$.}
  \label{fig:RIDC4-BE_start}
\end{figure}

During most of a RIDC calculation, multiple solution levels are
marched in a pipe using multiple computing CPUs/GPUs. However, the
computing nodes in the RIDC algorithm cannot start simultaneously:
each must wait for the previous level to compute sufficient $\eta$
values before they can be marched in a pipeline fashion.  By carefully
controlling the start-up of a RIDC method, one can minimize the amount
of memory that is required to march the nodes in a pipeline fashion.
In our implementation, the order in which computations are performed
during start-up is illustrated in Figure~\ref{fig:RIDC4-BE_start} for a
fourth order RIDC constructed with first order IMEX predictors and
correctors.  The $j$\textsuperscript{th} processor (running the
$j\textsuperscript{th}$ correction) must initially wait for $j(j+1)/2$
steps, e.g., node 2 has to wait 3 steps before starting.
There are also idle computing threads at the end of the computation,
since the predictor and lower level correctors will reach $t_N = b$
earlier than the last corrector.

An important notion to consider is ``restarts'', that is, instead of
computing all the way to the final time, we compute on some smaller
time interval to time $t_\star$, and use the most accurate solution to
restart the RIDC computation at $t=t_\star$.  In practice, restarting
improves the stability of the semi implicit RIDC scheme, and could
lower the error constant of the overall method, but at the cost of
decreasing the speedup due to the additional cost of starting the RIDC
algorithm multiple times.  

Additionally, one cannot increase the order of RIDC indefinitely as
(i) it is not practical (when would one ever want a 16th order
method?) and (ii) the Runge phenomenon \cite{Runge1901}, which arises
from using equi-spaced interpolation points, will eventually cause the
scheme to become unstable.  In practice, 8th and 12th order RIDC
methods using double precision do not suffer from the Runge phenomenon.

Lastly, there is another family of parallel time integrators, known as
parareal methods \cite{madayturinici02}, that is actively being
researched \cite{parareal-ornl,EmmettMinion11}.  These methods fall
into the class of ``parallel across the method'' algorithms, where the
entire time domain is split across multiple nodes, a coarse operator
is run in serial, followed by a parallel correction update.  We
encourage users to more carefully consider parareal if the small scale
parallelism offered by RIDC is not sufficient.

\section{Numerical Examples}
\label{sec:benchmarks}
Advection--reaction--diffusion equations have been widely used to
model chemical processes across many disciplines.  Here, we present
two numerical examples: an advection--diffusion and a
reaction--diffusion equation, to validate the order of accuracy of the
RIDC4-FBE scheme, and the speedup obtained in the parallel OpenMP
framework, and the OpenMP--CUDA hybrid framework. In each example, the
stiff term is chosen as the diffusion operator, $f^S(t,y)=y_{xx}$.
Applying a centered finite difference operator to approximate
$\partial_{xx}$ reduces each RIDC/ARK step to a series of decoupled
linear system solves.  The matrices are pre-factored into their QR
components so that each linear solve is reduced to a matrix--vector
multiplication and a back solve operation.

The computations presented were performed on a stand alone server
containing a quad core AMD Phenom X4 9950 2.6Ghz processor with four
Nvidia Tesla GPU C1060 cards (960 total GPU cores). (Superior speedups
will be observed with the newer Nvidia Tesla M2090 cards that are
rated at 665 Gflops at double precision, compared with the legacy
M1060 cards that are rated at 78 Gflops at double precision.)  The ARK
schemes are coded using plain C++ with (i) a homegrown linear algebra
library and (ii) the CUBLAS 4.0 library \cite{cuda}.  The RIDC4-FBE is
coded in C++ with (i) OpenMP and the homegrown linear algebra library,
and (ii) OpenMP and the CUBLAS 4.0 library.
%
Some important subtleties for creating a hybrid OpenMP -- CUDA RIDC
code are: (i) we can control which GPU card is used for a linear
solve by calling the {\tt cudaSetDevice()} function, and (ii) in using
``{\tt \#pragma for}'' loop to spawn individual threads for each
prediction/correction loop, we have to utilize static scheduling.


We note that to illustrate the effectiveness of parallel time
integrators in our numerical examples, we chose 1D problems so that by
taking a fine spatial resolution, the temporal discretization error
would dominate the spatial error.  Solutions to higher dimensional
solutions are practical with the newer available Fermi/Keppler cards
which have more onboard memory.  

\subsection{Advection-Diffusion}
\label{sec:ad}
\begin{figure}[htbp]
  \begin{center}
  \subfloat[Convergence study: error versus number of time steps]{
    \includegraphics[width=0.45\textwidth]{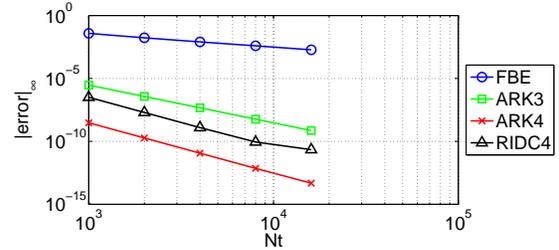}
    \label{fig:ad_conv}}
  \end{center}
  \subfloat[error versus wall clock time]{
    \includegraphics[width=0.45\textwidth]{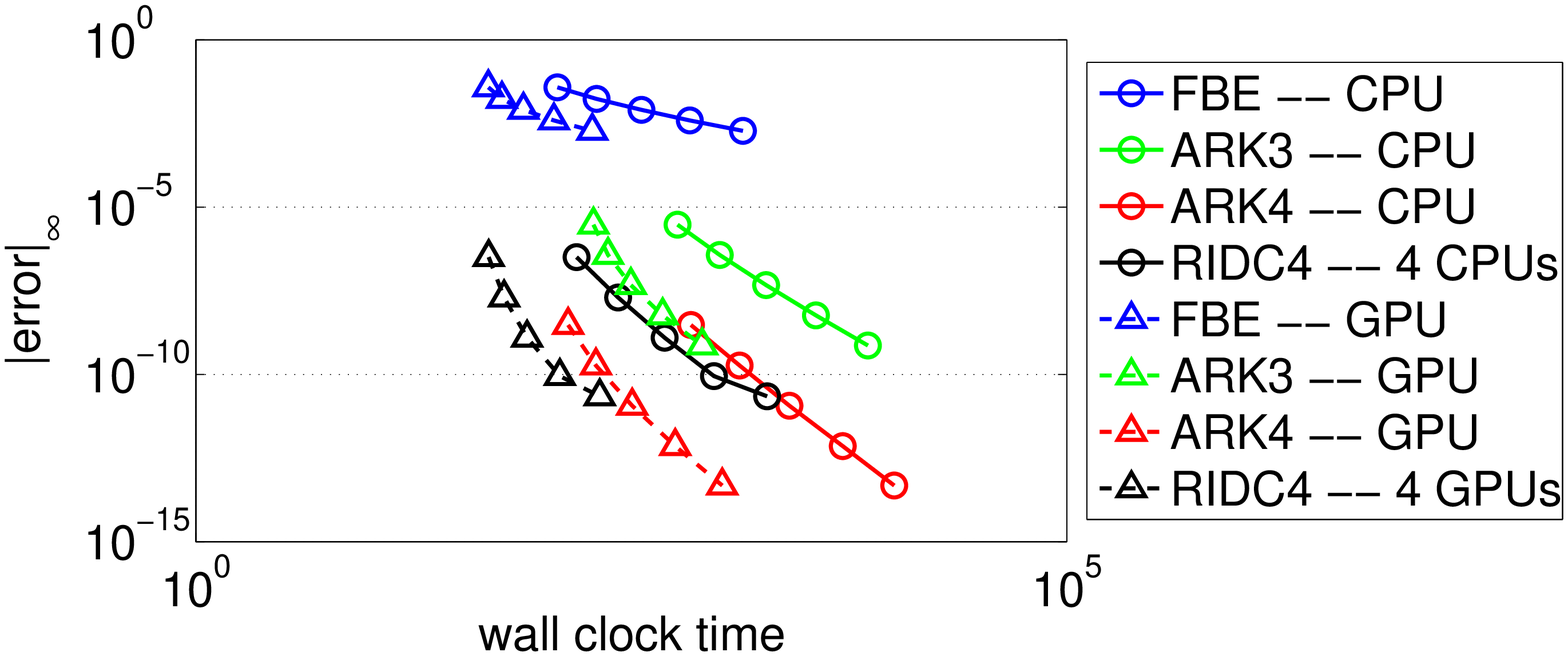}
    \label{fig:ad_timings}}
  \caption{(a) This standard ``error versus step size'' convergence
    study for RIDC4-FBE (with 10 restarts) and the various ARK methods
    presented in section~\ref{sec:ark}.  All schemes achieve their
    designed orders of accuracy.  Observe that the RIDC4-FBE error
    coefficient is much larger than that of the ARK4 scheme.  This is
    a small price to pay for the parallel speedup that can be
    obtained, as shown in (b).  Two observations should be made: (i)
    for all the schemes, our GPU implementation is approximately an
    order of magnitude faster than the CPU implementation, (ii) RIDC4
    (both the CPU and GPU implementations) compute a fourth order
    solution in the same wall clock as the FBE solution.}
    \label{fig:ad}
\end{figure}

We first consider the canonical advection-diffusion problem to show
that we can achieve designed orders of accuracy for our RIDC-FBE
algorithm.  The constant coefficient advection-diffusion equation,
\begin{align*}
  &u_t = cu_x + du_{xx}, \quad x \in [0,1], \quad t \in[0,40], \\
  &u(x,0) = 2 + \sin(2\pi x),
\end{align*}
with periodic boundary conditions, is discretized using the method of
lines methodology.  Specifically, the advection term is discretized
using upwind first order differences, and the diffusion term is
discretized using central differences.  The following system is then
recovered:
\begin{align*}
  u_t = Au + Du, \quad u(0) = \alpha,
\end{align*}
where the matrix $A$ approximates the advection operator, and the
matrix $D$ approximates the diffusion operator.  We choose the obvious
splitting, $f^N(t,u) = Au$, and $f^S(t,u) = Du$.  We take $c=0.1, d =
10^{-3}, \Delta x = \frac{1}{1000}$. First, we show in
Figure~\ref{fig:ad_conv} that ARK and RIDC4 achieve their designed
orders of accuracy.  Observe that the error coefficient for ARK4 is
several orders of magnitude smaller than that of RIDC4.  In
Figure~\ref{fig:ad_timings}, we instead plot the results from the same
numerical run, this time plotting error as a function of the wall
clock time.  Several observations can be made: (i) for all the
schemes, our GPU implementation is approximately an order of magnitude
faster than the CPU implementation, (ii) RIDC4 (both the CPU and GPU
implementations) compute a fourth order solution in the same wall
clock as the FBE solution, (iii) for a fixed wall clock time, RIDC4
(with 4 GPUs) computes a solution that is several orders of
magnitude more accurate than the solution computed using ARK4 (with 1
GPU).

\begin{figure}[htbp]
  \centering
  \includegraphics[width=0.4\textwidth]{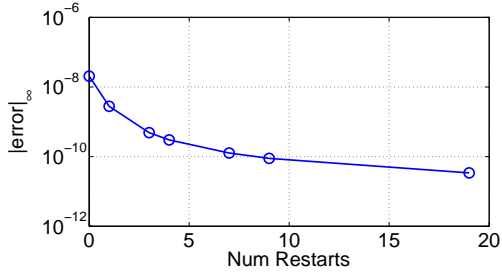}
  \caption{The error of RIDC4 schemes at the final time $T=40$
    decreases as the number of restarts is increased (for a fixed
    number of time steps, in this case, 4000 time steps). Each restart
    requires that the memory footprint be refilled before the
    cores/GPUs can be marched in a pipe.  }
  \label{fig:ad_restart}
\end{figure}

We also show in Figure~\ref{fig:ad_restart} the error of RIDC4 as a
function of restarts.  As expected, the error decreases as the number
of restarts is increased.  The penalty for each restart is having to
fill the memory footprint at each restart before marching the
cores/GPUs in a pipe.

Table~\ref{table:advection_speedup} summarizes the speedup that is
obtained when RIDC4 is computed using one, two and four CPUs, and when
RIDC4 is computed using one, two and four GPUs.  We appear to obtain
almost linear speedup, even with 10 restarts. This scaling is not
surprising since a bulk of the computational cost is due to the linear
solve and data transfer between host and GPU memory is not limited by
bandwidth for our example.
\begin{table}[htbp]
  \begin{center}
    \begin{tabular}{c|c}
      \# CPUs & Speedup\\
      \hline
      1 & 1.0 \\
      2 & 1.89\\
      4 & 3.81
    \end{tabular}\\
    \vspace*{0.1in}
    \begin{tabular}{c|c}
      \# CPUs \& GPUs & Speedup\\
      \hline
      1 & 1.0 \\
      2 & 1.97\\
      4 & 3.88
    \end{tabular}
  \end{center}  
  \caption{Speedup of RIDC for the advection--diffusion problem.}
  \label{table:advection_speedup}
\end{table}

The percentage of time that GPUs spend calling CUBLAS kernels are
summarized in the Table~\ref{table:cuda_profile}.  As the table
indicates, data transfer is a minimal component of our CUDA code.
\begin{table}[htbp]
  \begin{center}
    \begin{tabular}{c|c|c}
      kernel & calls & \% GPU time\\
      \hline
      trsv\_kernel & 5061 & 73.3\% \\
      gemv2N\_kernel\_ref & 11181 & 20.47\% \\
      gemv2T\_kernel\_ref & 5061 & 3.98\% \\
      axpy\_kernel\_ref & 24200 & 1.15\% \\
      memcpyHtoD & 8243 & 0.61\% \\
      memcpyDtoH & 5061 & 0.43\% 
    \end{tabular}
  \end{center}
  \caption{Profiling our GPU code for the advection--diffusion problem.}
  \label{table:cuda_profile}
\end{table}


\subsection{Viscous Burgers' Equation}
\begin{figure}[htbp]
  \centering
  \includegraphics[width=0.3\textwidth]{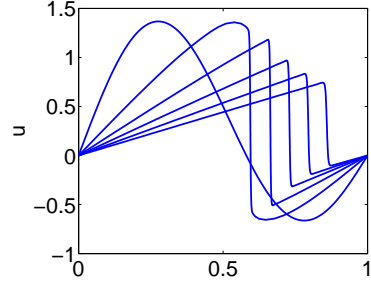}
  \caption{Solution to Burgers' equation, with $\epsilon=10^{-3}$ and
    $\Delta x = \frac{1}{1000}$.  Time snapshots at $t=0, 0.2, 0.4,
    0.6, 0.8$ and $1$ are shown.}
  \label{fig:burger_sol}
\end{figure}

We also consider the solution to viscous Burgers' equation,
\begin{align*}
  u_t + \frac12\left(u^2\right)_x  = \epsilon u_{xx},
  \quad (x,t)\in[0,1]\times[0,1],
\end{align*}
with initial and boundary conditions
\begin{align*}
u(0,t) = u(1,t) = 0, \quad u(x,0) = \sin{(2\pi x)}+\frac12\sin{(\pi x)}.
\end{align*}
The solution develops a layer that propagates to the right, as shown
in Figure~\ref{fig:burger_sol}.
 The diffusion term is again discretized using centered finite
 differences.  A numerical flux is used to approximate the advection
 operator,
\begin{align*}
  \frac12 ( (u_i^n)^2 )_x =
  \frac12\frac{f^n_{i+1/2} - f^n_{i-1/2}}{\dx},
\end{align*}
where
\begin{align*}
f_{i+1/2}^{n} = \frac12\left((u_{i+1}^n)^2 + (u_{i}^n)^2\right).
\end{align*}
Hence, 
the following system of equations is obtained,
\begin{align*}
  u_t = \mathcal{L}(u) + Du, 
\end{align*}
where the operator $\mathcal{L}(u)$ approximates the hyperbolic term
using the numerical flux, and the matrix $D$ approximates the
diffusion operator.  We choose the splitting $f^N(t,u) =
\mathcal{L}(u)$ and $f^S(t,u) = Du$, and take $\epsilon=10^{-3}$ and
$\Delta x = \frac{1}{1000}$.  No restarts are used for this simulation.

\begin{figure}[htbp]
   \subfloat[Convergence study: error versus number of time steps]{
    \includegraphics[width=0.45\textwidth]{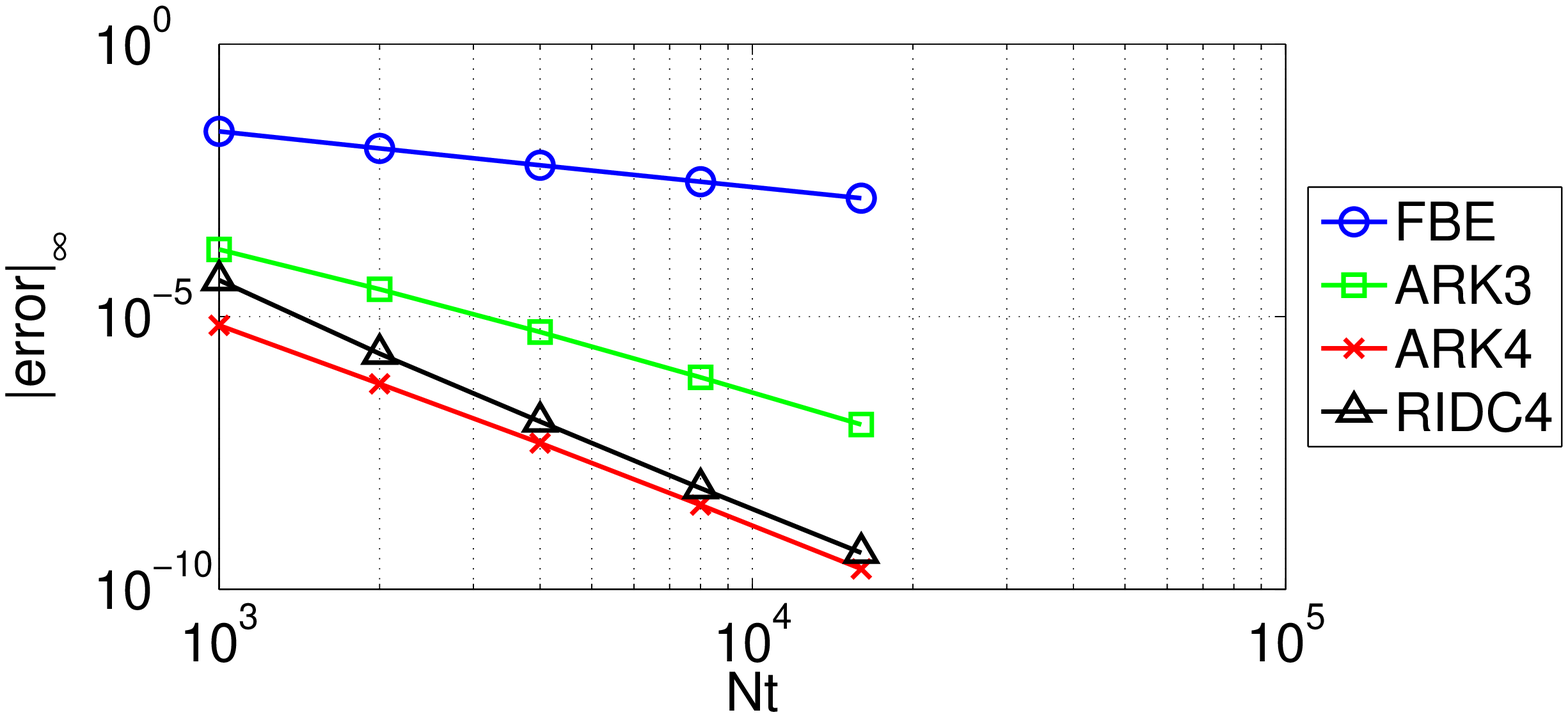}
  \label{fig:burger_conv}}\\
   \subfloat[error versus wall clock time]{
    \includegraphics[width=0.45\textwidth]{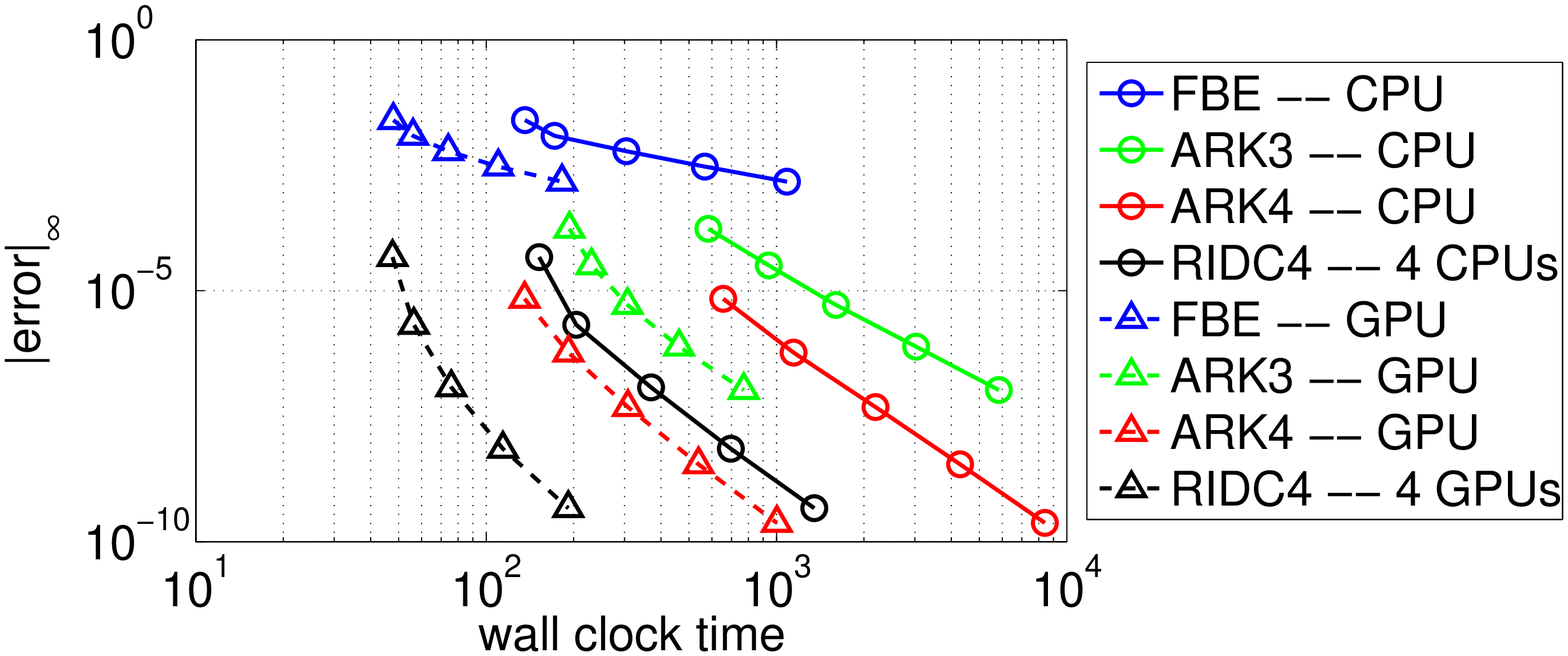}
  \label{fig:burger_timings}}
  \caption{In (a), we show that the ARK schemes and our RIDC4-FBE
    scheme achieve the designed orders of accuracy.  The plot in (b)
    shows the error as a function of wall clock time.  Two
    observations should be made: (i) for all the schemes, our GPU
    implementation is approximately an order of magnitude faster than
    the CPU implementation.}
  \label{fig:burger}
\end{figure}

The same numerical results as the previous advection--diffusion
example are observed in Figure~\ref{fig:burger}.  In plot (a), the
RIDC scheme achieves it's designed order of accuracy.  In plot (b), we
show that our RIDC implementations (both the CPU and GPU versions)
obtain a fourth order solution in the same wall clock time as a first
order semi-implicit FBE solution.  The RIDC implementations with
multiple CPU/GPU resources also achieve comparable errors to a fourth
order ARK scheme in approximately one tenth the time.  

Table~\ref{table:burger_speedup} summarizes the speedup that is
obtained when RIDC4 is computed using one, two and four CPUs, and when
RIDC4 is computed using one, two and four GPUs.
\begin{table}[htbp]
  \begin{center}
    \begin{tabular}{c|c}
      \# CPUs & Speedup\\
      \hline
      1 & 1.0 \\
      2 & 1.94\\
      4 & 3.94
    \end{tabular}\\
    \vspace*{0.1in}
    \begin{tabular}{c|c}
      \# CPUs \& GPUs & Speedup\\
      \hline
      1 & 1.0 \\
      2 & 1.98\\
      4 & 3.95
    \end{tabular}
  \end{center}  
  \caption{Speedup of RIDC for Burgers' equation.}
  \label{table:burger_speedup}
\end{table}

\section{Conclusions}
\label{sec:conclusion}
In this paper, we further developed RIDC algorithms to generate a
family of high order semi-implicit parallel integrators.  The analysis
related to convergence is a simple extension from previous work, and
the numerical experiments demonstrate that the fourth order RIDC-FBE
algorithm achieves its designed order of accuracy.  Additionally, we
showed that our semi-implicit RIDC algorithm harnessed the
computational potential of four GPUs by utilizing OpenMP coupled
with with the CUBLAS library.  This semi-implicit RIDC algorithm can
potentially be coupled with existing legacy parallel spatial codes.
Work is on-going to explore a hybrid MPI--OpenMP--CUDA algorithm for
more heterogeneous architectures.

\section*{Acknowledgments}
This work was supported by AFRL and AFOSR under contract and grants
FA9550-07-0092 and FA9550-07-0144 and NSF grant number DMS-0934568.
We also wish to acknowledge the support of the Michigan State
University High Performance Computing Center and the Institute for
Cyber Enabled Research.

\bibliography{ridc}
\bibliographystyle{plain}

\end{document}